\documentclass[twocolumn,showpacs,preprintnumbers,amsmath,amssymb]{revtex4}

\usepackage[dvips]{color}
\usepackage{graphicx}
\begin{document}
%%%%%%%%%%%%%%%%%%%%%%%%%%%%%%%%%%%%%%%%%%%%%%%%%%%%%%%%%%%%%%%%%%%%%%
\title{Detecting Majorana nature of neutrinos in muon and tau decay}
\pacs{14.80.Cp,%Non-standard-model Higgs bosons
14.60.Pq,%Neutrino mass and mixing
12.60.Fr,%Extensions of electroweak Higgs sector
13.35.Bv%Decays of muons
}
\preprint{IC/2008/082} 
%%%%%%%%%%%%%%%%%%%%%%%%%%%%%%%%%%%%%%%%%%%%%%%%%%%%%%%%%%%%%%%%%%%%%%
\author{Takeshi Fukuyama}
\email{fukuyama@se.ritsumei.ac.jp} \affiliation{Department of
Physics and R-GIRO, Ritsumeikan University, Kusatsu, Shiga,
525-8577, Japan} \affiliation{SISSA, via Beirut 2-4 I-34014, Trieste, Italy}
%%%%%%%%%%%%%%%%%%%%%%%%%%%%%%%%%%%%%%%%%%%%%%%%%%%%%%%%%%%%%%%%%%%%%%
\author{Koji Tsumura}
\email{ktsumura@ictp.it} \affiliation{The Abdus Salam ICTP of
UNESCO and IAEA, Strada Costiera 11, 34014 Trieste, Italy}
%%%%%%%%%%%%%%%%%%%%%%%%%%%%%%%%%%%%%%%%%%%%%%%%%%%%%%%%%%%%%%%%%%%%%%
\begin{abstract}
The Majorana nature of neutrinos can be detected 
by the precise measurement of muon decay. 
This possibility comes from the presence of charged Higgs boson 
interaction for Majorana neutrinos. 
We study the effects of the neutrino Yukawa interaction via 
charged Higgs bosons in muon decay processes 
such as $\mu\to e\nu\overline{\nu}$ and $\mu\to e\gamma$. 
The Higgs triplet model with small vacuum expectation value is of special importance 
whose neutrino Yukawa coupling can affect significantly muon decays.
External neutrino lines in the Feynman diagrams of $\mu\to e\nu\overline{\nu}$ 
can be crossed because of its Majorana nature. 
This fact provides the interference contribution between 
the W boson exchange diagram and that of charged Higgs boson, 
which may be detectable in near future experiments. 
\end{abstract}
%%%%%%%%%%%%%%%%%%%%%%%%%%%%%%%%%%%%%%%%%%%%%%%%%%%%%%%%%%%%%%%%%%%%%%
\maketitle
%%%%%%%%%%%%%%%%%%%%%%%%%%%%%%%%%%%%%%%%%%%%%%%%%%%%%%%%%%%%%%%%%%%%%%

We have obtained many definite information on neutrino over the last decade, showing manifestly new physics beyond the standard model (SM).
There are still undetermined parameters $\theta_{13}$, absolute mass values 
and leptonic CP phase in neutrino sector. 
Determinations of mass parameters are not the end of the story. 
The smallness of neutrino masses implies new origin of mass scale which 
may be related to the Majorana nature of neutrinos. 
The question of whether neutrino is Majorana or Dirac particles is one of the most important issues in particle physics. 

The direct detection of Majorana nature of neutrino has been tried in neutrinoless double beta decay~\cite{genius,cuore,moon}.
New type of experiments using the interference of neutrinos emitted from high Rydberg 
atom is also on going~\cite{Yoshimura}. 
These attempts use the rare decay processes. 
On the other hand, the direct detection of the Majorana nature 
by precise measurement of muon decay $\mu\to e\nu\overline{\nu}$ was also studied 
in the model with V+A currents~\cite{Takasugi1,Takasugi2}.
In the seesaw framework\cite{Seesaw}, these effects are characterized 
by the mixing between neutrinos and anti-neutrino which is determined 
by a ratio $m_D^{}/M_R^{}$  where 
$m_D^{}$ is the Dirac mass and $M_R$ is the Majorana mass for neutrinos. 
The presence of the V+A current can affect 
the muon decay rate, but the deviation distinguishing Majorana from Dirac 
neutrino are suppressed by ${\mathcal O}\left((m_D^{}/M_R^{})^2\right)$. 
Even if we assume the scale of right-handed neutrinos 
to be TeV region, i.e., $M_R\sim \mathcal{O}(\text{TeV})$ 
with $m_\nu\sim\mathcal{O}(\text{eV})$~\cite{Lazarides,Goh}, the deviation is of ${\mathcal O}\left(10^{-12}\right)$ at most. However, present experimental accuracies of muon decay is $10^{-3}\%$ level~\cite{Chitwood:2007pa}. 
So the detection is still out of the scope in near future experiments in their framework. 

In this letter, we show that Majorana nature of neutrinos may be detectable by precise measurements of muon decay and the other decay processes. 
This is realized by the minimal extension of SM, including only the $SU(2)_L$ scalar triplet additionally (Higgs triplet model; HTM~\cite{HTM}).
A detectability of the Majorana nature of neutrinos is 
discussed by using the interference effect of muon decay amplitude 
between W-boson exchange diagrams and that of charged Higgs boson in the HTM. 
The deviation of the muon decay rate from the SM prediction can reach 
several times $10^{-4}\%$. It can be detectable at further precise 
measurements of muon decay. 
Our method also gives the possibility to measure the effective neutrino mass 
for tau leptons in its leptonic decay.

Before studying the case for the HTM, 
we discuss the effects of the charged Higgs from more general standpoint, 
which is necessary to assure that the deviation is indeed due to the interference of Majorana neutrinos.
That is, we argue the effects due to the large neutrino Yukawa coupling 
in the two-Higgs-doublet model (THDM)~\cite{HHG} first, and 
then in the HTM. 
These effects would be measured at future low energy experiments for relatively 
small vacuum expectation values of extra Higgs bosons $(v_\nu,v_\Delta^{}\ll v)$. 
It is shown that the deviation of $\Gamma(\mu\to e\nu\overline{\nu})$ from SM can become detectable level 
only in the HTM because of the Majorana nature of neutrinos.

%%%%%%%%%%%%%%%%%%%%%%%%%%%%%%%%%%%%%%%%%%%%%%%%%%%%%%%%%%%%%%%%%%%%%%
In the SM, the weak interaction Lagrangian is written by
\begin{align}
{\mathcal L}_{W^\pm}^{} &= -\frac{g}{\sqrt2}\,U_{\ell i}^{}\,
\overline{\ell_L^{}}\gamma^\alpha \nu_i W^-_\alpha+\text{H.c.}
\end{align}
where $L, \nu$ are lepton doublet and neutrino fields in each mass
diagonal bases, and $U_{\ell i}$ is the neutrino mixing matrix.
The muon decay rate %with massive Dirac or Majorana neutrinos 
is calculated as 
\begin{align}
\Gamma_{\mu\to e \nu{\overline \nu}}^\text{SM}
=\frac{G_F^2m_\mu^5}{192\pi^3}\,f\left(\frac{m_e^2}{m_\mu^2}\right)(1+R.C.)
\times\left(1+\frac{3m_\mu^2}{5m_W^2}\right). \label{Eq:MuDecay}
\end{align}
Here $G_F$ is defined by
\begin{equation}
G_F\equiv \frac{g^2}{4\sqrt{2}M_W^2}
\end{equation}
with the universal weak coupling constant $g$. The function is 
$f(x)=1-8x+8x^3-x^4-12x^2\ln x$, and $R.C.$ represents radiative 
corrections which is given by \cite{Stuart}
\begin{align}
R.C.=&\frac{\alpha}{2\pi}\left(\frac{25}{4}-\pi^2\right)\left[1+\frac{\alpha}{\pi}\left(\frac{2}{3}\ln\frac{m_\mu}{m_e}-3.7\right) \right.\nonumber \\
&\left.+\left(\frac{\alpha}{\pi}\right)^2\left(\frac{4}{9}\ln^2\frac{m_\mu}{m_e}-2.0\ln\frac{m_\mu}{m_e}+C\right)\right].
\end{align}
As we will show soon, the signal of Majorana nature can be $O(10^{-6})$, 
and we have written in Eq.(\ref{Eq:MuDecay}) up to this precision order.

Let us begin with the THDM, where we consider an additional Higgs doublet 
which interacts with Dirac neutrinos. All other fermions only couples to the SM 
Higgs doublet. 
There are three neutral Higgs bosons 
and a pair of charged Higgs boson $H^\pm$. 
In order to generate small neutrino masses, the vacuum 
expectation value $(v_\nu)$  of the extra Higgs doublet can be taken 
as small $v_\nu/v \ll 1$. 
Such small vacuum expectation value predicts a very light scalar 
which is experimentally allowed because of the highly suppressed couplings 
with gauge bosons and charged fermions. 
Its detailed phenomenology is discussed in ~\cite{Gabriel:2006ns}. 
Neutrino masses are obtained by the diagonalization of 
$m_\nu=y_\nu v_\nu/\sqrt2$. In this setup, the neutrino Yukawa coupling 
can be taken to be large. 
In the THDM, the muon decay rate is corrected by the 
contribution of charged Higgs boson through the charged 
lepton Yukawa interaction~\cite{Krawczyk}. This effect is 
suppressed by a factor $v_\nu/v$. We here consider the extent 
of the large neutrino Yukawa coupling. 
Neutrino Yukawa interaction with charged Higgs bosons are given by
\begin{align}
{\mathcal L}_{H^\pm}^{} &= -\frac{\sqrt2{m_\nu}_i}{v_\nu^{}}\,U_{\ell
i}^{}\, \overline{\ell_L^{}}\nu_iH^-+\text{H.c.}
\end{align}
where $\ell$ and $\nu$ are charged lepton and neutrino fields 
in mass diagonal basis of charged leptons, ${m_\nu}_i$ represents neutrino masses, and 
$U_{\ell i}(i=1$--$3)$ is the neutrino mixing matrix. 
The effect of the neutrino Yukawa interaction in muon decay is calculated as 
\begin{align}
\delta\Gamma_{\mu\to e \nu{\overline \nu}}^\text{THDM}&=
\frac{G_F^2m_\mu^5}{192\pi^3}\,f\left(\frac{m_e^2}{m_\mu^2}\right)
\left(\frac{v}{v_\nu}\right)^4
\frac{\langle m_\nu^2\rangle_\mu\langle m_\nu^2\rangle_e}{4m_{H^\pm}^4},
\label{Eq:H-Contrib}
\end{align}
where $\langle m_\nu^2\rangle_\ell=\sum_i{m_\nu}_i^2|U_{\ell i}|^2$. 
In the above formula, we neglect neutrino masses but keep 
its Yukawa interaction, i.e., ${m_\nu}_i/m_\ell\ll {m_\nu}_i/v_\nu (\ell=e,\mu)$. 

%%%%%%%%%%%%%%%%%%%%%%%%%%%%%%%%%%%%%%%%%%%%%%%%%%%%%%%%%%%%%%%%%%%%%%

The muon lifetime has been precisely measured 
$\tau_\mu =\left(2.197019\pm0.000021\right)\times 10^{-6}$s
~\cite{Chitwood:2007pa}.
Once charged Higgs boson is observed, and its mass is 
determined at the LHC, the deviation of the muon decay rate 
from the SM prediction can constrain $v_\nu$.
The search for charged Higgs bosons has been examined at LEP 
via its pair production. It gives the lower limit on the mass of 
charged Higgs boson $m_{H^\pm}^{}\gtrsim 79.3$ GeV\cite{Particle}. 
Requiring the small deviation less than $10^{-3}$\%,  
the lower limit of $v_\nu$ would be obtained as 
\begin{align}
v_\nu\gtrsim 2\text{eV}\left(\frac{m_\nu}{0.05\text{eV}}\right)
\left(\frac{100\text{GeV}}{m_H^\pm}\right).
\end{align}

%%%%%%%%%%%%%%%%%%%%%%%%%%%%%%%%%%%%%%%%%%%%%%%%%%%%%%%%%%%%%%%%%%%%%%

The large neutrino Yukawa coupling induces rare muon 
decay $\mu\to e\gamma$. Its decay branching fraction is calculated as 
\begin{align}
\text{Br}(\mu\to e\gamma)^\text{THDM}
=\frac{\alpha_\text{EM}^{}}{24\pi}\left(\frac{v}{v_\nu}\right)^4
\frac{\left|{m_\nu}_j^2U_{ej}U_{\mu j}\right|^2}{m_{H^\pm}^4}.
\end{align}
Taking into account the experimental upper limit 
$\text{Br}(\mu\to e\gamma)^\text{exp} <1.2\times 10^{-11}$~\cite{MuEGammaExp}, 
we obtain the slightly stronger lower bound on $v_\nu$, 
\begin{align}
v_\nu\gtrsim 6\text{eV}\left(\frac{m_\nu}{0.05\text{eV}}\right)\left(\frac{100\text{GeV}}{m_H^\pm}\right).
\label{MuE}
\end{align}
Applying the bound in Eq.~\eqref{MuE}, 
the possible deviation of $\Gamma(\mu\to e\nu\overline{\nu})$ is   
only several times $10^{-8}$ which will not be measured in near future. 

%%%%%%%%%%%%%%%%%%%%%%%%%%%%%%%%%%%%%%%%%%%%%%%%%%%%%%%%%%%%%%%%%%%%%%

Next we discuss the effect of Majorana neutrinos on muon decay 
in the framework of the HTM. 
We introduce an $SU(2)_L$ triplet Higgs boson $\Delta$ in addition to the SM.  
The origin of neutrino masses can be adapted by the interactions term,
\begin{align}
{\mathcal L}_\text{HTM}^{} &= \overline{L^c}h_M\,i\tau_2\Delta
L+\text{H.c.}
\end{align}
Here neutrinos are required to be Majorana particles. The matrix
$h_M^{}$ is coupling strength and $\tau_i(i=1$--$3)$ denote 
the Pauli matrices.
The triplet Higgs boson field with hypercharge $Y=2$ 
can be parameterized by
\begin{align}
\Delta=\begin{pmatrix}\Delta^+/\sqrt2&\Delta^{++}\\
\frac{v_\Delta^{}}{\sqrt2}+\Delta^0&-\Delta^+/\sqrt2\end{pmatrix},
\end{align}
where $v_\Delta^{}$ is the vacuum expectation value of triplet Higgs boson. 
Mass eigenvalues of neutrinos are determined by diagonalization of
$m_\nu^{}=\sqrt2h_M^{}v_\Delta^{}$. 
There is a tree level contribution to rho parameter from the 
triplet vacuum expectation value as $\rho\thickapprox 1-2v_\Delta^2/v^2$.
LEP precision result gives an upper 
limit $v_\Delta^{}\lesssim 5$ GeV. 
There is no stringent bound from $b\to s\gamma$ 
on the charged Higgs boson mass because the triplet Higgs boson 
does not couple to quarks. 

\begin{figure}[tb]
\centering
\includegraphics[width=8cm]{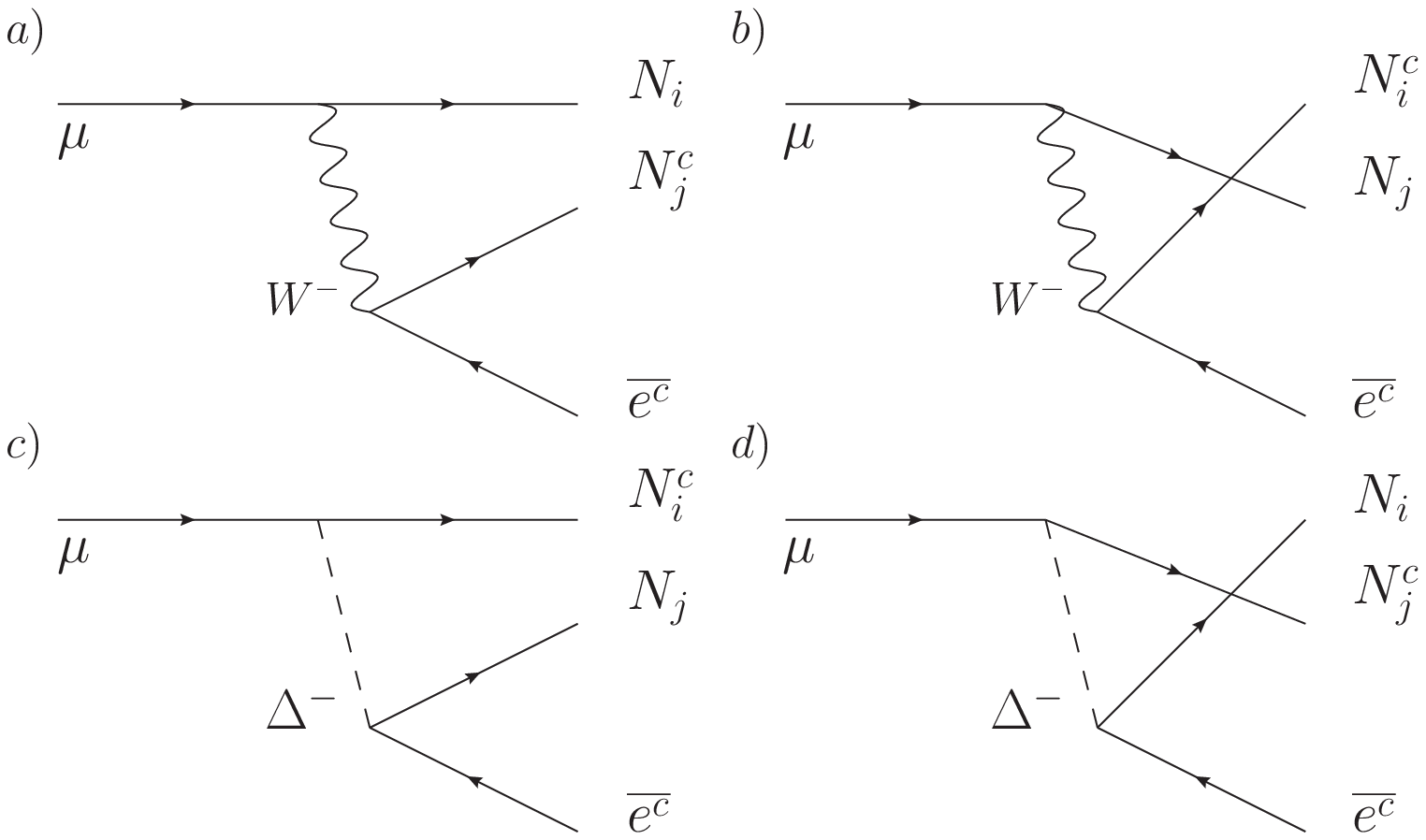}
\caption{The Feynman diagrams for the muon decay $\mu^-\to
e^-N{\overline N}$.} \label{FIG:MuDecay}
\end{figure}

The Yukawa interaction of the singly and the doubly charged Higgs boson
is written by
\begin{align}
{\mathcal L}_{\Delta}^{} =& -\frac{{m_\nu}_i}{v_\Delta^{}}\,U_{\ell
i}^{}\, \overline{\ell_L^{}}N_i^c\Delta^-\nonumber \\
&-\frac{{m_\nu}_i}{\sqrt2 v_\Delta^{}}\,U_{\ell i}^{}U_{\ell' i}^*\, \overline{\ell_L^{}}{{\ell'}_L}_i^c\Delta^{--}
+\text{H.c.},
\end{align}
where $N_i$ represent Majorana neutrinos which satisfy conditions $N_i=N_i^c=C\overline{N_i}^T$. Therefore Majorana fields can contract not only 
with fermions but also with anti-fermions. 
In FIG.~\ref{FIG:MuDecay}, we depict the Feynman
diagrams for the muon decay in the HTM.
The contribution to the muon decay from the singly charged 
Higgs boson is calculated as 
\begin{align}
&\delta\Gamma_{\mu\to e N{\overline N}}^\text{HTM}=
\frac{G_F^2m_\mu^5}{192\pi^3}\,f\left(\frac{m_e^2}{m_\mu^2}\right)
\nonumber \\
&\times\left[2\left(\frac{v}{v_\Delta^{}}\right)^2
\frac{\left|\langle m_\nu\rangle_{\mu e}\right|^2}{m_{\Delta^\pm}^2}
+\left(\frac{v}{v_\Delta^{}}\right)^4
\frac{\langle m_\nu^2\rangle_\mu\langle m_\nu^2\rangle_e}{m_{\Delta^\pm}^4}
\right],
\label{interference}
\end{align}
where the effective mass of neutrino is defined as $\langle m_\nu\rangle_{\mu e}=\sum_j{m_\nu}_jU_{\mu j}^*U_{e j}$. 
The first term in Eq.~\eqref{interference} 
comes from the interference between Fig.(1a) and (1d).
The presence of interference term is a consequence of the Majorana nature 
of neutrinos. Again we note that in the above formula we neglect neutrino 
masses but keep its Yukawa coupling. The effective neutrino masses should be 
understood by the ratio of the Yukawa coupling and 
the triplet vacuum expectation value. 

Let us estimate the magnitude of these effects. 
The most stringent constraint on the triplet Yukawa coupling 
comes from $\mu\to ee{\bar e}$ through the tree level contribution due to 
the doubly charged Higgs boson~\cite{Gunion}. The branching ratio for this decay 
is given by 
\begin{align}
\text{Br}(\mu\to ee{\bar e})^\text{HTM}=\frac18
\left(\frac{v}{v_\Delta^{}}\right)^4
\frac{\left|\langle m_\nu\rangle_{\mu e}\langle m_\nu\rangle_{ee}\right|^2}{m_{\Delta^{\pm\pm}}^4}.
\end{align}
The experimental bound 
$\text{Br}(\mu\to ee{\bar e})^\text{exp}<1.0\times 10^{-12}$~\cite{Mu3eExp} 
gives the upper limit for the neutrino Yukawa coupling, 
which is translated into the lower bound 
on $v_\Delta^{}$
\begin{align}
v_\Delta^{}\gtrsim 70\text{eV}\left(\frac{m_\nu}{0.05\text{eV}}\right)\left(\frac{100\text{GeV}}{m_{\Delta^{\pm\pm}}^{}}\right)^2.
\end{align}
In order to avoid the large one loop contribution 
to rho parameter from the triplet Higgs bosons, we take their masses 
to be degenerate~\cite{rhoHTM}. 
Under these conditions, the decay branching ratio of 
$\mu\to e\gamma$ is suppressed enough. 
The deviation from the SM muon decay rate can reach to several times 
$10^{-4}\%$. The difference between 
the effect and the current experimental accuracy is only a few factor. 
It might be accessible by further precise measurement of the muon decay. 
Unfortunately, muon decay is used as a precision measurement of 
Fermi coupling constant $(G_F)$, and this deviation is renormalized in $G_F$~\cite{Marciano}. So we need 
an independent determination of $G_F$ for final decision of Majorana nature. 
%However, the data of electroweak precision measurements are expected to be accumulated from various experiments in the very near future and our proposal may be one of the most important themes of the precision measurements.

%%%%%%%%%%%%%%%%%%%%%%%%%%%%%%%%%%%%%%%%%%%%%%%%%%%%%%%%%%%%%%%%%%%%%%

We comment on the detectability of the effective neutrino masses 
for tau leptons. The decay formulae for tau decay $\tau\to\ell N\overline{N}
,(\ell=e,\mu)$  are easily obtained by substituting $\mu\to\tau$ and $e\to\ell$ 
in the preceding formula. Experimental accuracies of such tau decays are 
not so good, ${\mathcal O}(10^{-2})$~\cite{Hayasaka}. However, 
the upper bounds for the tau associated neutrino Yukawa couplings in the HTM 
from the lepton flavor violating decays $\tau\to \ell\ell'\ell''$ 
are also loose. In fact, upper bounds for these branching fractions are order 
$10^{-4}$ smaller than that of $\mu\to ee{\bar e}$. 
Therefore, we can expect larger deviation from the SM predictions 
on this process. 
The tau leptonic decays may be useful to distinguish the Dirac 
and Majorana nature of neutrinos even if the effect of neutrino Yukawa 
coupling is renormalized.

We have discussed detectability of Majorana nature in muon decay 
$\mu\to e\nu\overline{\nu}$. 
The effect of the large neutrino Yukawa coupling on muon decays 
are studied in the THDM and the HTM.  
In the THDM with small vacuum expectation value of extra Higgs boson, 
the size of neutrino Yukawa coupling is restricted by the upper limit 
of $\mu\to e\gamma$. Under the constraint, we have evaluated 
the possible deviation of decay rate for $\mu\to e\nu\overline{\nu}$. 
Its effect is not so large which is out of the scope in near future 
experiments.
In the HTM, the larger values of neutrino Yukawa couplings are strongly 
bounded by non observation of $\mu\to ee{\bar e}$.
Despite of stringent limits, we have found 
that muon decay rate can deviate to accessible level.
This possibility comes from the additional indifference contribution 
between the W-boson and the charged Higgs boson mediation diagrams 
through the Majorana nature of neutrinos. 
We have shown that the effective masses for tau leptons can also be 
observed in tau leptonic decays.  

{\bf Acknowledgments}~~~\\[2mm]
We would like to thank M.~Yoshimura, K.~Lynch, K.~Hayasaka, M.~Tanaka, H.~Nishiura, H.~Sugiyama and S.~Kanemura for very useful conversations. 
T.F. is grateful to S.Petcov for his hospitality at SISSA.
The work of T.F. is supported in part by the grant-in-Aid 
for Scientific Research from the Ministry of Education, 
Science and Culture of Japan (No. 20540282). 

%%%%%%%%%%%%%%%%%%%%%%%%%%%%%%%%%%%%%%%%%%%%%%%%%%%%%%%%%%%%%%%%%%%%%

%%%%%%%%%%%%%%%%%%%%%%%%%%%%%%%%%%%%%%%%%%%%%%%%%%%%%%%%%%%%%%%%%%%%%%
\end{document}